\documentclass{article}

\usepackage{arxiv}

\usepackage[utf8]{inputenc} 
\usepackage[T1]{fontenc}    
\usepackage{hyperref}       
\usepackage{url}            
\usepackage{booktabs}       
\usepackage{amsfonts}       
\usepackage{nicefrac}       
\usepackage{microtype}      
\usepackage{lipsum}		
\usepackage{graphicx}
\usepackage{natbib}
\usepackage{doi}
\usepackage{amsmath}
\usepackage{amssymb}
\usepackage{amsbsy}
\usepackage{amsthm}
\usepackage{caption,subcaption}

\newcommand{\wt}{\widetilde}
\newcommand{\wh}{\widehat}

\DeclareMathOperator*{\diag}{Diag\,}

\newtheorem{theorem}{Theorem}

\newtheorem{lemma}{Lemma}

\title{Continuous Optimization for Offline Change Point Detection and Estimation}

\author{ Hans Reimann\\
	Department of Mathematics\\
	University of Potsdam\\
	Potsdam, BRANDENBURG, GERMANY \\
	\texttt{hans.reimann@uni-potsdam.de} \\
	\And
	\href{https://orcid.org/0000-0003-2868-9420}{\includegraphics[scale=0.06]{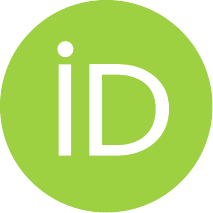}\hspace{1mm} Sarat Moka} \\
	School of~Mathematics and Statistics\\ 
    University of New South Wales\\
    Sydney, NSW, AUSTRALIA \\
	\texttt{s.moka@unsw.edu.au} \\
 	\And
	\href{https://orcid.org/0000-0001-5342-7559}{\includegraphics[scale=0.06]{orcid.pdf}\hspace{1mm} Georgy Sofronov} \\
	School of Mathematical and Physical Sciences\\
    Macquarie University\\ 
    Sydney, NSW, AUSTRALIA\\
	\texttt{georgy.sofronov@mq.edu.au} \\
}

\renewcommand{\shorttitle}{Offline Change Point Detection and Estimation}

\hypersetup{
pdftitle={Continuous Optimization for Offline Change Point Detection and Estimation},
}

\begin{document}
\maketitle

\begin{abstract}
This work explores use of novel advances in best subset selection for regression modelling via continuous optimization for offline change point detection and estimation in univariate Gaussian data sequences. The approach exploits reformulating the normal mean multiple change point model into a regularized statistical inverse problem enforcing sparsity.
After introducing the problem statement, criteria and previous investigations via Lasso-regularization, the recently developed framework of continuous optimization for best subset selection (COMBSS) is briefly introduced and related to the problem at hand. Supervised and unsupervised perspectives are explored with the latter testing different approaches for the choice of regularization penalty parameters via the discrepancy principle and a confidence bound. The main result is an adaptation and evaluation of the COMBSS approach for offline normal mean multiple change-point detection via experimental results on simulated data for different choices of regularisation parameters. Results and future directions are discussed.  
\end{abstract}


\section{Introduction}
\label{sec:intro}

Change point detection and estimation are an incredibly diverse and widely scattered field in applied and mathematical statistics, with a large variety of applications. To provide a high-level intuition, change point detection may be understood as a signal processing tool for identifying abrupt changes in the generative parameters of a data sequence.
While a strong line of work in change point detection is well established with Page's pioneering work (see \cite{page1954continuous}) and rigorous results by \cite{chernoff1964estimating}, 
\cite{lorden1971procedures} and \cite{sen1975tests}, many aspects of this problem are open and the general understanding of good solutions depends strongly on the problem at hand \cite{niu2016multiple}, \cite{truong2020selective}, and \cite{Ma2020}. Among the open research questions, the simultaneous detection of multiple change points in large data sets is of major interest.

Taking a machine learning and data scientific motivated approach, in this paper, we explore the applicability of recent advances in best subset selection of covariates in linear regression proposed by \cite{moka2024combss}. This method, a continuous optimization approach for best subset selection, claims to offer faster performance compared to existing exhaustive search methods, while maintaining comparable accuracy. Furthermore, it achieves superior accuracy relative to methods of similar speed, thus embodying highly desirable properties for addressing the subset selection problem in offline change point detection and estimation. Some of the existing methods including \cite{huang2005detection}, 
\cite{rinaldo2009properties}
and \cite{qian2016stepwise} utilize Lasso-regularisation which provides convex relaxation of the underlying sparsity problem. However, such Lasso-regularization approaches do not directly focus on optimizing the actual discrete constrained best subset selection problem at hand. 

In this paper, we focus on the popular normal mean multiple change point model assuming univariate, independent Gaussian random variable with a constant variance. This simple yet challenging model has applications in several fields including genomics and econometrics. Further, detection of change points can be reduced to detecting change in mean in adapted sequences \cite{niu2016multiple} such as detection of change in slope reducing to detection of mean in the difference sequence. The challenge in detection and estimation of an unknown number of change points in a piece-wise identical distributed sequence of observations resembles a problem in choice of regularization penalty parameter in practice.
   
We will explore two different methods and compare their performance assuming knowledge on variance of the noise. This work formulates offline change point detection as a best subset selection problem and it allows us to explore recent advances in this domain. Our goal is to provide a proof of concept as well as a stepping stone for future investigations. In particular, we employ the novel advances in continuous optimization for best subset selection in the given context to provide a new and modern perspective in approaching the change point detection problem via modern machine learning methods. We believe that the employed framework of simultaneous offline multiple change point detection as a 
sparse linear regression problem has high potential in utilizing the broad established toolbox of both areas. 

The remaining paper is organized as follows. In Section \ref{sec:theo}, we formulate the normal mean multiple change point model as a best subset selection problem in linear regression and then discuss the key results of  \cite{moka2024combss}. Our method for change point detection problem is presented in Section~\ref{sec:meth}. In Section~\ref{sec:complexity}, linear time running time complexity of the proposed method is presented. In Section~\ref{sec:exp} the simulation experimental designs and the corresponding results  are presented. Finally, a discussion on the results and future directions for further studies are presented in Section~\ref{sec:conc}. 

\section{Theoretical Background}
\label{sec:theo}
In this section, we first provide formulation of offline multiple change point detection as a subset selection problem in linear regression via boundary conditions on sparsity of covariates  (see \cite{rinaldo2009properties} and \cite{niu2016multiple}). Then, we provide the gradient based continuous optimization introduced of \cite{moka2024combss} towards solving the subset selection problem. 

\subsection{Change Point Detection via Subset Selection}

Let $Y = (Y_1, \dots, Y_n)$ be a sequence of independent random variables with the distribution of $Y_i$ denoted as~$F_i$. A change point is then a point in the set of indices of the sequence $\tau \in \{2, \dots, n\}$ such that  $F_{\tau-1}\neq F_\tau$. The central aim in change point detection lies in estimating the location of an instance $\tau$. In multiple change point detection, the goal is to estimate both the locations of the change points and their total number, denoted as $K$. 

Consider the multiple mean change point detection problem with $K$ change points with their locations denoted by $\tau_1, \dots, \tau_K$ where $\tau_k \in\{2, \dots, n\}$ and $\tau_k < \tau_{k+1}$ for each $k \in \{1,\ldots,K-1\}$. Note that the case where $K=n-1$ (i.e., an instance of change with every next sequence member) has generally very little insights as the interest usually lies in detecting piece-wise independent and identically distributed ($iid$) random variables in between instances of change. On the other extreme, the case where $K \leq 1$ (i.e., whether a single change point is present or not) is of great interest in that it quantifies whether a full $iid$ assumption on a sequence of data is justified. For the normal mean multiple change point model, which is central in this work, we assume $K > 1$ and take $Y_i\sim\mathcal{N}(\mu_i, \sigma^2)$ for each $i \in \{1, \dots, n\}$ with mean $\mu_i$ and common variance $\sigma^2$. A change point $\tau_k$ now refers to an unknown location in the sequence $Y$ with $\mu_{\tau_k-1}\neq\mu_{\tau_k}$. In addition, we make the usual assumption that the total number of change points $K$ is unknown. Consequently, the normal mean multiple change point model can be understood as a sequence $Y$ generated by $K+1$ piece-wise constant signals $\mu_i$ and zero-mean $iid$ noise sequence $\varepsilon_1, \dots, \varepsilon_n \sim \mathcal{N}(0,\sigma^2)$, so that 
\[
Y_i=\mu_i+\varepsilon_i, \quad i \in \{1,2,\ldots,n\}.
\]
The central challenge then translates to estimating the mean sequence 
\[
\mu_1=\ldots=\mu_{\tau_1-1}\neq\mu_{\tau_1}=\mu_{\tau_1+1}=\ldots=\mu_{\tau_k-1}\neq\mu_{\tau_k}=\ldots=\mu_{\tau_K-1}\neq\mu_{\tau_K}=\ldots=\mu_n.
\]
Besides, we introduce an artificial change point at $\tau_0=1$ whenever the initial mean $\mu_1\neq0$. 

 We now briefly discuss two major criteria in change point detection based on  signal-to-noise ratio and sparsity; refer to, e.g., \cite{niu2016multiple} for more details. Both these criteria often provide the main assumptions and quantities of interest in investigation and analysis. In order for the instances of mean change  to be insightful and identifiable, they need to have relevant degree of distinction from one another which is measured in shift between means relative to the noise. The corresponding parameter is the minimal difference in signal standardized by the noise standard deviation, formally defined as,
 \[
 \delta=\underset{i\in\{2,3,\ldots,n\}}{\min}\frac{|\mu_i-\mu_{i-1}|}{\sigma},
 \]
 which is hereafter simply referred to as signal-to-noise ratio. Additionally, two instances of change in too close proximity do not allow for any valuable inference as there is too little data in between. This motivates control of the minimal distance between instances of change, which is formally defined as 
 \[
 L =\underset{k\in\{1,2,\ldots,K\}}{\min}\{\tau_k-\tau_{k-1}\}.
 \]
 The assumption on sparsity hereby also implicitly forms a boundary condition of distance of the first and last instance of change from the respective start and end of the data sequence. Further following \cite{niu2016multiple}, both these parameters are combined into the central quantity of signal strength $S=\delta^2 L$. While establishing the weakest possible condition on $S$ for full recovery of all change points is an open problem, results by \cite{arias2005near} show that $S \geq 2\log n$ could be a necessary condition for full recovery of multiple change points.  Additionally, \cite{rinaldo2009properties} and \cite{qian2016stepwise} showed for their fused Lasso regression approach that the consistency $\mathbb{P}(\hat{\mathbf{\tau}}=\mathbf{\tau}) \rightarrow1$ as $n\to\infty$ with $\mathbf{\tau} =(\tau_1, \dots, \tau_K) \in \mathbb{R}^K$ denoting the vector of change point locations, where  $\hat\tau$ being the estimator of $\tau$. 
 However, this requires a fairly strong assumption of $\delta^2 \gg \log n$, which may not be applicable in practice as indicated in \cite{niu2016multiple}.
 
 The above mentioned linear regression approach to change point detection is best understood in terms of the statistical inverse problem of recovering the mean vector $\mu$ from the data sequence $y$, a realization of the random vector $Y$. While a $L^1$-distance for tackling the problem may be reasonable under considerations of robustness, we want to apply a $L^2$-norm instead for desirable properties on differentiability. This results in the ordinary least squares formulation $\|y - \mu\|^2_2$. Via introducing an artificial design matrix $X$ and parameter vector $\beta$, we can further transform the problem to a typical statistical inverse problem. In particular, let $\beta_i = \mu_i - \mu_{i-1}$ for $i \in \{1, 2, \dots, n\}$ with $\mu_0 = 0$ and $X$ be a $n\times n$ lower-triangular matrix with all the non-zero elements being $1$.
Then, we obtain $\mu = X\beta$ and $Y = X\beta+\varepsilon$.
The resulting statistical inverse problem is then
 \begin{equation*}
     \underset{\mu\in\mathbb{R}^n}{\min}\|y - \mu\|^2_2=\underset{\beta\in\mathbb{R}^n}{\min}\|y - X\beta\|^2_2,
 \end{equation*}
which is the ordinary least squares approach for estimating the parameter vector $\beta$ in linear regression. By construction, the piece-wise constant structure of the mean vector $\mu$ translates to $0$ entries in $\beta$ as 
 \[
 \mu_i = \mu_{i-1}\quad \text{if and only if}\quad \beta_i =0, 
 \]
for all $i \in\{2, 3, \dots, n\}$. This motivates the use of modern high-dimensional sparse regression methods for recovery of $\mu$, the central connection exploited in this work. 
 
With the given definition of $\beta$, we observe that $\beta_i\neq0$ for $i=\tau_k$ and $k\in\{0,1,\ldots K\}$, possibly including the artificial change point $\tau_0$.
That is, imposing sparsity in instances of change translates to controlling the total number of non-zero entries of $\beta$. Hence, we obtain
 \begin{align}
 \label{eqn:min_s}
         \underset{\beta\in\mathbb{R}^n}{\min}\frac{1}{n}\|y - X\beta\|^2_2 \quad      \mathrm{subject}\ \mathrm{to}\, \, \|\beta\|_0 := \sum_{i=1}^n \mathbb{I}(\beta_i\neq0)\leq K,
 \end{align}
 where $\mathbb I(\cdot)$ denotes the usual indicator function.
This problem is a reformulation of the normal mean change point detection to an ordinary least squares regression problem with the sparsity induced by the $L^0$-constraint $\|\beta\|_0 \leq K$. This is a well studied highly non-convex best subset selection problem in linear regression; refer to \cite{mueller2010selection} and \cite{hui2017joint} for details.

Among the existing methods for solving \eqref{eqn:min_s}, the Lasso regression is a popular convex relaxation of the subset selection problem where a $L^0$-norm $\|\beta\|_0$ is replaced by a $L^1$-norm $\|\beta\|_1$. In particular, the convex relaxation of the Lasso regularization is given by
 \begin{align*}
         \underset{\beta\in\mathbb{R}^n}{\min}\frac{1}{n}\|y - X\beta\|^2_2 + \lambda \|\beta\|_1,
 \end{align*}
where the parameter $\lambda$ in the penalty allows us to control the sparsity in the solution. Application of the Lasso regularization to the normal mean change point detection was studied in \cite{huang2005detection} and \cite{harchaoui2010multiple} with additional constraints for a fused Lasso penalty explored in \cite{tibshirani2008spatial} and further work in \cite{rinaldo2009properties} and \cite{qian2016stepwise}. 
While this change yields desired properties in fast computation and estimation, it does not generally guarantee to address the best subset problem as above; refer, e.g., \cite{zhu2020polynomial} and \cite{hazimeh2020fast}. 
However, the main insight from all these studies is the necessity of tuning $\lambda$ in penalty \cite{friedman2007pathwise}, e.g. via cross-validation as heuristic for parameter choice.

Two very recent works by \cite{wang2020univariate} and \cite{verzelen2023optimal} support the sketched approach and show minimax optimality of their respective estimators. Additionally, in emphasizing the value of post-processing of detected instances of change, they provide a valuable future line of work. Both these works focus on deriving lower bounds for the described multiple change point model, however, with some adaptations and extensions, i.e., a more general sub-Gaussian noise term. Each then introduces an estimator based on the best subset selection approach with additional focus on the estimation of the total number of changes. Yet in both cases, estimators constructed based on \eqref{eqn:min_s} are shown to reach minimax optimal rates with high computational efficiency. The gradient based estimator described in this work aims to fall in line with these results and provide a solution via continuous optimization similar to the Lasso approach.

\subsection{Continuous Optimization for Subset Selection}
We now introduce the key idea of \cite{moka2024combss} that enables a continuous, gradient based optimization for best subset selection. Towards this, recall the optimization problem~\eqref{eqn:min_s} and consider a vector $s\in\{0,1\}^n$ with 
\begin{equation*}
    s_i=\begin{cases}
		1, & \text{if $i=\tau_k$ for some $k\in\{0,1,\ldots K\}$},\\
        0, & \text{otherwise},
		 \end{cases}
\end{equation*}
that is, the indices of non-zero elements $s$ correspond to the location of the change points. Hence, $|s| = \sum_{i=1}^n s_i = K$ is equivalent to $\|\beta\|_0 = K$. With this notation, for any $s \in \{0, 1\}^n$, let $X_{[s]}$ be the matrix of size $n\times |s|$ obtained from $X$ 
by keeping only the columns with indices $j$ where $s_j=1$. 
Then, for any $K\in\{1,2,\ldots p\}$, the exact best subset problem \eqref{eqn:min_s} can be restated as
\begin{align}
\label{eqn:min_Xs}
\underset{s\in\{0,1\}^n}{\min}\frac{1}{n}\|y - X_{[s]}\wh{\beta}_{[s]}\|^2_2,
         \quad&\mathrm{subject}\ \mathrm{to} \, \, |s| \leq K,
\end{align}
where $\hat{\beta}_{[s]}$ is the low-dimensional ordinary least squares solution utilizing $X_{[s]}$ instead of the full model design matrix $X$ for estimation.

Even for a given $K$, the resulting problem in \eqref{eqn:min_Xs} is in general non-deterministic polynomial-time hard (NP-hard) as stated in \cite{natarajan1995sparse}. Exact methods are only feasible for small values of $n$. A more recent approach via mixed integer optimization as in \cite{bertsimas2016best}, although faster than the exact method, remains fairly slow for practical application \cite{hazimeh2020fast}. The Lasso regularization, mentioned above, is a reasonable relaxation regarding feasibility concerns, however, it does not recover the best subset as in \eqref{eqn:min_s} or \eqref{eqn:min_Xs}, as stated before recalling \cite{hazimeh2020fast} and \cite{zhu2020polynomial}. As a novel approach, continuous optimization method towards best subset selection, COMBSS, aims to combine computational feasibility and accuracy in enabling modern gradient based continuous optimization methods for the best subset selection problem; see COMBSS with {\sf SubsetMapV1} in \cite{moka2024combss} for details. 

The central idea of COMBSS is to relax the binary vector $s\in\{0,1\}^n$ in \eqref{eqn:min_Xs}, which takes values at the corners of the hypercube $[0, 1]^n$, to a vector $t\in[0,1]^n$ evolving on the entire hypercube. 
For each such $t$, let 
\[
M_t = X_t^\top X_t + n\, (I - \diag(t\odot t)),
\]
with $X_t=X\mathrm{Diag}(t)$, where $I$ is the identity matrix, $\diag(v)$ is the diagonal matrix with $v$ as its diagonal, and $\odot$ denotes the Hadamard (or, element-wise) product. Further, let $\wt{\beta}_t$ be a solution of the linear equation (in terms of $u$), 
\begin{align}
\label{eqn:linear_eqn}
M_t u = X_t^\top y.
\end{align}
Then, the continuous Boolean relaxation of \eqref{eqn:min_Xs} is
\begin{align}\label{eqn:min_Xt}
         \underset{t\in[0,1]^n}{\min}\frac{1}{n}\|y-X_{t}\wt{\beta}_{t}\|^2_2,
         \quad \mathrm{subject}\,\, \mathrm{to} \sum_{i=1}^nt_i \leq K.
\end{align}
The construction of $\tilde{\beta}_t$ hereby guarantees equality of (\ref{eqn:min_Xt}) and \eqref{eqn:min_Xs} at the corner points of the hypercube $[0,1]^n$. Instead of solving this linear constrained problem, COMBSS aim to optimize its Lagrangian form as 
\begin{align}
\label{eqn:lagrange_Xt}
         \underset{t\in[0,1]^n}{\min} \frac{1}{n}\|y - X_{t}\wt{\beta}_{t}\|^2_2+\lambda\sum_{i=1}^n t_i,
\end{align}
where the regularization penalty parameter $\lambda$, analogous to $K$ in the exact problem, allows us to control the sparsity in the solution obtained. Further, the continuous relaxation inside the hypercube enables the desired smoothness in $t$ of the objective function $\|y - X_{t}\wt{\beta}_{t}\|^2_2$ and the Lagrangian function in \eqref{eqn:lagrange_Xt}. COMBSS main insight lies in suggesting candidates of best subsets by utilizing this smoothness, without discrete constraints, while directly addressing the best subset selection unlike previous continuous optimization approaches. This way popular gradient based optimization methods in machine learning,  such as Adam optimizer, can easily be applied to optimize \eqref{eqn:lagrange_Xt}. Further, via a transformation from $[0,1]^n$ to $\mathbb{R}^n$, COMBSS transforms the box-constrained optimization \eqref{eqn:lagrange_Xt} to an equivalent unconstrained optimization so that the continuous optimizer does not face any boundary issues. The final point $t$ for each $\lambda$ is mapped to an approximate solution $s$ of \eqref{eqn:min_Xs} by taking $s_i = 1$ if $t_i$ is close to $1$, otherwise $s_i = 0$, for all $i$. 

The algorithm and exact mathematical framework with full details of the procedure as well as the respective guarantees for smoothness are given in \cite{moka2024combss}. The key enabling feature for computational feasibility lies in efficiently solving the linear equation \eqref{eqn:linear_eqn}. Further, there is software made available for both Python and R to implement the method. Our experimental results emphasize the ability of COMBSS to recover best subsets in both low and high dimensions. To summarize, by providing smooth solution paths in the continuous relaxation of the exact best subset selection problem, COMBSS has enabled a new and promising line of work. Among the benefactors may well be the best subset selection problem in simultaneous normal mean multiple change point detection.  

\section{Methodology}
\label{sec:meth}

The central idea of this work is eminent: Normal mean multiple change point detection and estimation in recovering the piece-wise constant mean vector $\mu$ is equivalent to finding the best subset of covariates via sparsity conditions imposed on the artificial design matrix $X$ and parameter vector $\beta$ with $\mu=X\beta$ in~\eqref{eqn:min_s}. This optimization problem is equivalent to the reduced lower-dimensional discrete optimization problem in~\eqref{eqn:min_Xs}. Solving the best subset selection using an exact method, such as {\em leaps-and-bounds}, is only computationally feasible when $n$ is less than $31$ \cite{furnival2000regressions}. However, the novel results of \cite{moka2024combss} enable approaching it via continuous optimization  for solving \eqref{eqn:lagrange_Xt}. Furthermore, the regularisation penalty parameter $\lambda$ enables us to control the total number of change points. Thus, we create a grid of $\lambda$ values and execute COMBSS for each $\lambda$ to produce a vector $t_\lambda^* \in [0, 1]^n$, which is further mapped to a binary vector $s_\lambda^* \in \{0,1\}^n$.  This binary vector $s_\lambda^*$ has $1$'s at locations corresponding to the change points. So, using $s_\lambda^*$, we compute $\wh \beta_{[s_\lambda^*]}$ with the non-zero elements corresponding to the shift in the mean values at the change points.


Taking a supervised perspective, where we assume that the number of change points $K$ is known, the problem then translates into finding a suitable $\lambda$ over a grid corresponding to $K$ change points. This is straightforward, e.g., via a simple grid search or interval-halving algorithm. 
However, the more practical problem is to take an unsupervised perspective assuming that $K$ is unknown. This turns the challenge at hand into a typical regularisation penalty parameter choice for statistical inverse problems. While we do have a better understanding of the respective penalty parameter $\lambda$ compared to the previously introduced Lasso approaches, it still leaves us with a similar conclusion to \cite{rinaldo2009properties} in the unsupervised case -- it is an important open problem. 

To address this issue of penalty parameter choice, we intend to employ established methods for regularization choice in the context of inverse problems. In particular, we investigate the discrepancy principle and an adaptation of the discrepancy principle based on a lower confidence bound. 
Let $\beta_\lambda$ denote the parameter vector $\beta$ obtained for a given choice of $\lambda$. Following \cite{richter2021inverse}, the discrepancy principle for statistical inverse problems aims to choose $\lambda$ such that the error of $\|y-X\beta_\lambda\|^2_2$ aligns with the expected model error, $\mathbb{E}[\|Y-X\beta\|^2_2]$. However, an analytical evaluation of this expectation is only feasible in specific cases for the data sequence $y$. Given the data sequence as introduced via the normal mean multiple change points and constant noise variance $\sigma^2$, we can standardize the inverse problem and find $\beta_\lambda$ such that it recovers $\|\frac{1}{\sigma}(y-X\beta_\lambda)\|^2_2=n$, as under the true model $\|\frac{1}{\sigma}(y-X\beta)\|^2_2\sim\chi^2(n)$ with expectation equal to the degrees of freedom. Again, the basic idea hereby is that the regularization penalty is chosen so that the average or expected level of error given the estimated parameter vector should resemble that of a true parameter vector given the model. For a much more sophisticated and thorough analysis of the discrepancy principle for statistical inverse problems see \cite{blanchard2012discrepancy}. Adaptation to a confidence bound extends this comparison via using a $\chi^2$-quantile instead of the expectation, to control the effect of the model error in the inverse problem. This then leads to the choice of a regularization penalty that is at least as restrictive as with the discrepancy principle. The intuition here is that of a one-sided composite hypothesis of the regularisation penalty $\lambda$ that recovers the parameter vector $\beta$ in $H_0:\ \lambda\leq\lambda_0$ vs $H_1:\ \lambda>\lambda_0$ for $\lambda_0$ the penalty parameter recovering the true model. The $\chi^2$-distribution of the standardized regularization again provides a decision threshold. The regularisation penalty is chosen such that the supposedly plausible error is maximal within the confidence bound $\underset{\lambda>0}{\max}\|\frac{1}{\sigma}(y-X\beta_\lambda)\|^2_2\leq\chi^2_{1-\alpha}(n)$ for $\alpha\in(0,1)$. In essence this means increasing $\lambda$ and thus decreasing $K$, right until $H_0$ needs to be rejected. The basic idea then is that under the true parameter vector the accumulated model error  is plausible  up to the threshold $\chi^2_{1-\alpha}(n)$. As noted, this accommodates at least as much error as the discrepancy principle and therefore may allow for more restrictive choices of regularisation penalty and smaller values of $K$. The main limitation of the discrepancy principle and its adaptation is that they require knowledge of the noise variance of the sequence. whereas they do not require a constant variance assumption.
Solutions via estimating the variance from the sequence itself are difficult, but could be implemented via pooled variance estimators. Via an iterative scheme, repeated estimation of the pooled covariance and change point detection may provide ways of implementation of the discrepancy principle even for unknown covariance. However, this is likely to lead to problems in dependence of the estimated variance and the analysis of the expected error after standardization. Accordingly, the main insight for the following experiments lies in a simple litmus test of whether and to what extent information about the variance helps in the choice of the regularisation penalty.


\section{Time Complexity}
\label{sec:complexity}
We now show that our implementation of COMBSS exhibits $O(n)$ running time complexity. In the general COMBSS approach of \cite{moka2024combss}, for optimizing \eqref{eqn:lagrange_Xt}, computational complexity of the gradients of the objective function is dictated by computations of solving linear equations of the form $M_t u = b$. The matrix $M_t$ can be shown to be dense positive-definite for $t \in (0,1)^n$ and thus, exact gradient computation can be expensive with a time complexity of $O(n^3)$. When a conjugate gradient method is used for approximately solve such a linear equation, the general COMBSS exhibits $O(n^2)$ complexity. 

It is important to note that in the general setting, the design matrix $X$ is part of the data. However, in our case, $X$ has an artificial design of being lower triangular matrix with the non-zero elements equal to~$1$. We now take advantage of this fact to modify the COMBSS algorithm to achieve a time complexity of $O(n)$.
Towards this, Lemma~\ref{lem:XX_inv} establishes that the inverse of $X^\top X$ is tridiagonal. 
\begin{lemma}
\label{lem:XX_inv}
Consider the lower triangular matrix $X$ with the non-zero elements being $1$. Then, $A = (X^\top X)^{-1}$ is a tridiagonal matrix with the principle diagonal being $(1, 2, 2, \dots, 2)$ (i.e., first element is $1$ and all other elements are equal to $2$) and both the sub-diagonals being $(-1, -1, \dots, -1)$ (i.e., all elements equal to $-1$).
\end{lemma}
We can easily establish Lemma~\ref{lem:XX_inv} using a proof by induction. Start with the basic case at $n = 2$ and assume that it is true for $n$. Then, we can show that this is true for $n+1$ via using the Banachiewicz inversion lemma \cite{tian2005schur}. 

\begin{lemma}[Woodbury Matrix Identity]
\label{lem:woodbury}
For any conformable matrices $A$, $U$, $C$ and $V$,
\[
(U A V + C)^{-1} = C^{-1} - C^{-1} U (A^{-1} + V C^{-1} U)^{-1} V C^{-1}.
\]
\end{lemma}
Refer to \cite{woodbury1950inverting} for a proof of Lemma~\ref{lem:woodbury}.
Note that with $T_t = \diag(t)$ and $D_t = d(I - T_t^2)$,
we can write 
$M_t = \left( T_t X^\top X T_t + D_t\right)/n$.
Further, as a consequence of Lemma~\ref{lem:woodbury}, we have the following result
\begin{theorem}
\label{thm:Lt-inverse}
Let $\wt M_t = (X^\top X)^{-1} + T_{t} D_t^{-1} T_t$.
Then, for the lower triangular matrix $X$, 
$\wt M_t$ is a tridiagonal matrix, and 
\begin{align}
\label{eqn:Lt_inv_expand}
    M_t^{-1} = n\, D_t^{-1} - n\, D_t^{-1} T_t \wt M_t^{-1} T_t D_t^{-1}, \quad t \in [0, 1)^n.
\end{align}
\end{theorem}
Proof of Theorem~\ref{thm:Lt-inverse} is straightforward: Since $(X^\top X)^{-1}$ is tridiagonal given in Lemma~\ref{lem:XX_inv} and $T_{t} D_t^{-1} T_t$ is diagonal, $\wt M_t$ must be tridiagonal. Selecting $A = X^\top X$, $C = D_t$, and $U = V = T_t$ in Lemma~\ref{lem:woodbury}, we get \eqref{eqn:Lt_inv_expand}.

Recall that we want to compute the expressions of the form $M_t^{-1}b$. Since $D_t$ is diagonal, computing expressions of the form $D_t^{-1} v$ has $O(n)$ complexity. This means, the bottleneck is the complexity of computing expressions of the form $\wt M_t^{-1} v$. Since $\wt M_t$ is a tridiagonal matrix, this operation is easy to execute using the popular method of the {\em Thomas algorithm} or {\em tridiagonal matrix algorithm} which is known to have $O(n)$ complexity; see, e.g., \cite{lee2011tridiagonal} and \cite{higham2002accuracy}. In conclusion, the computational complexity of solving a linear equation of the form $M_t u = b$ is $O(n)$ only.  \cite{moka2024combss} have shown that the COMBSS method approaches an $\epsilon$-stationary point in $O(1/\epsilon^2)$ iterations (independent of $n$). Thus, our implementation has $O(n)$ time complexity.

\section{Empirical Evaluation Results}
\label{sec:exp}

By construction and with the extensive simulation results in \cite{moka2024combss}, the proposed approach appears reliable in best subset selection in linear regression. The experimental simulation study in this paper aims to investigate the two levels of knowledge. First, there is the supervised level with knowledge about the number of change points $K$. How well does our continuous optimization method recover the locations of change given $K$ is known? Second is the unsupervised level with knowledge about the variance of the noise. How well do the discrepancy principle and the confidence bound perform for our method in recovering the true number of change points as well as their locations? 
All experiments follow a basic set up reduced to its essentials. We conduct two sets of experiments for known and unknown number of change points. In each set, the first respective experiment scales signal-to-noise ratio, $\delta^2$, with a fixed distance between change points. The second experiment then scales minimal distance between change points, $L$, for fixed signal-to-noise ratio. We set $\sigma^2=1$ throughout all experiments only adjusting the difference in mean values to set signal-to-noise ratio. We conduct $100$ Monte Carlo simulations of the noise for each experiment and value of the scaling quantities $\delta$ and $L$. Following \cite{aminikhanghahi2017survey} and \cite{truong2020selective}, the metrics for evaluation will be the F1-score and the Hausdorff distance metric. The F1-score hereby serves as a measure of accuracy combining sensitivity and precision with a tolerance level of $\tau_k\pm L/20$ to allow for negligible inaccuracies. The Hausdorff distance metric serves as evaluation of robustness of the detection procedure in measuring the largest difference between a change point and its closest estimated counterpart. For better comparison under scaling minimal distance, we also standardize the Hausdorff metric by $L$. The evaluation criteria are averaged over all Monte-Carlo simulations for a concrete value of $\delta$ or $L$. 
All experiments are conducted in R (version 4.2.2).

\subsection{Experiment A - Known Number of Change Points}
For each Monte-Carlo simulation, the regularization penalty $\lambda$ for the proposed algorithm is chosen via a interval halving approach until it produces a result with the right number of change points $K$. In very rare instances, if the interval halving fails to produce a result with right number $K$ in a certain number of steps, the corresponding Monte-Carlo sample will be skipped in the aggregation of the Hausdorff metric. 

\subsubsection{Experiment A1 - Scaling $\delta^2$}

Set $n=150$ with a total of $4$ change points at $\tau=(31,61,91,121)$, so $L=30$. The mean sequence has a simple staircase form with difference in mean at the change points $\mu_{\tau_k}-\mu_{\tau_k-1}$ for $k\in\{1,2,3,4\}$ ranging from $0.25$ to $4$ with step-size $\triangle\delta=0.25$ and $\mu_{\tau_0}=0$, so it should not be detected. The necessary condition in \cite{arias2005near} is therefore $\delta_\mathrm{crit}\geq\sqrt{\frac{2\log(150)}{30}}\approx0.58$ and is surpassed after two steps of $\delta$.
\begin{figure}[h!] 
  \begin{subfigure}{0.5\linewidth}
    \centering
    \includegraphics[width=0.7\textwidth]{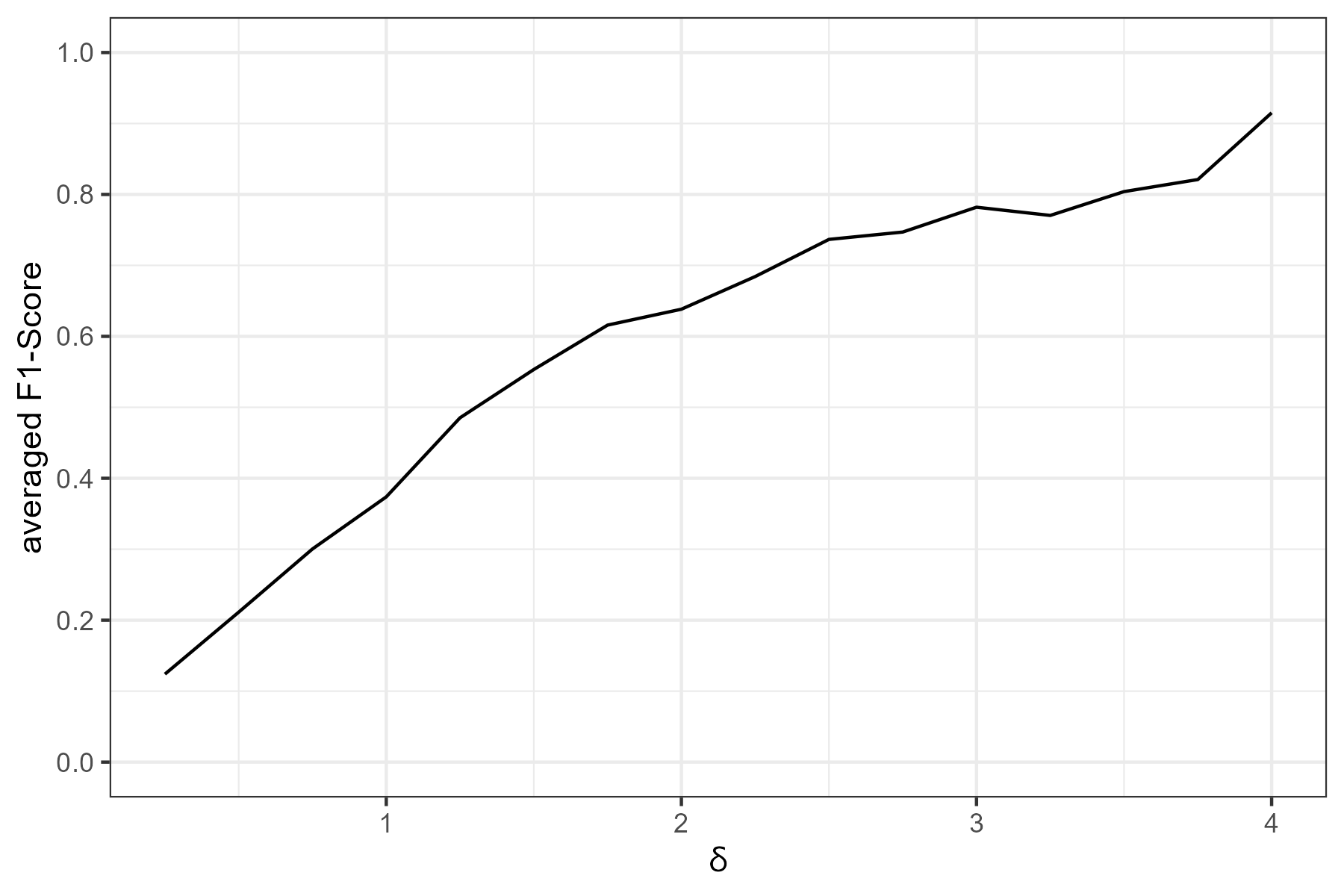}
    \caption{Scaling signal-to-noise ratio via $\delta$ and resulting\\
    F1-score. \label{fig: mu_f1}} 
    \end{subfigure}
  \begin{subfigure}{0.5\linewidth}
    \centering
    \includegraphics[width=0.7\textwidth]{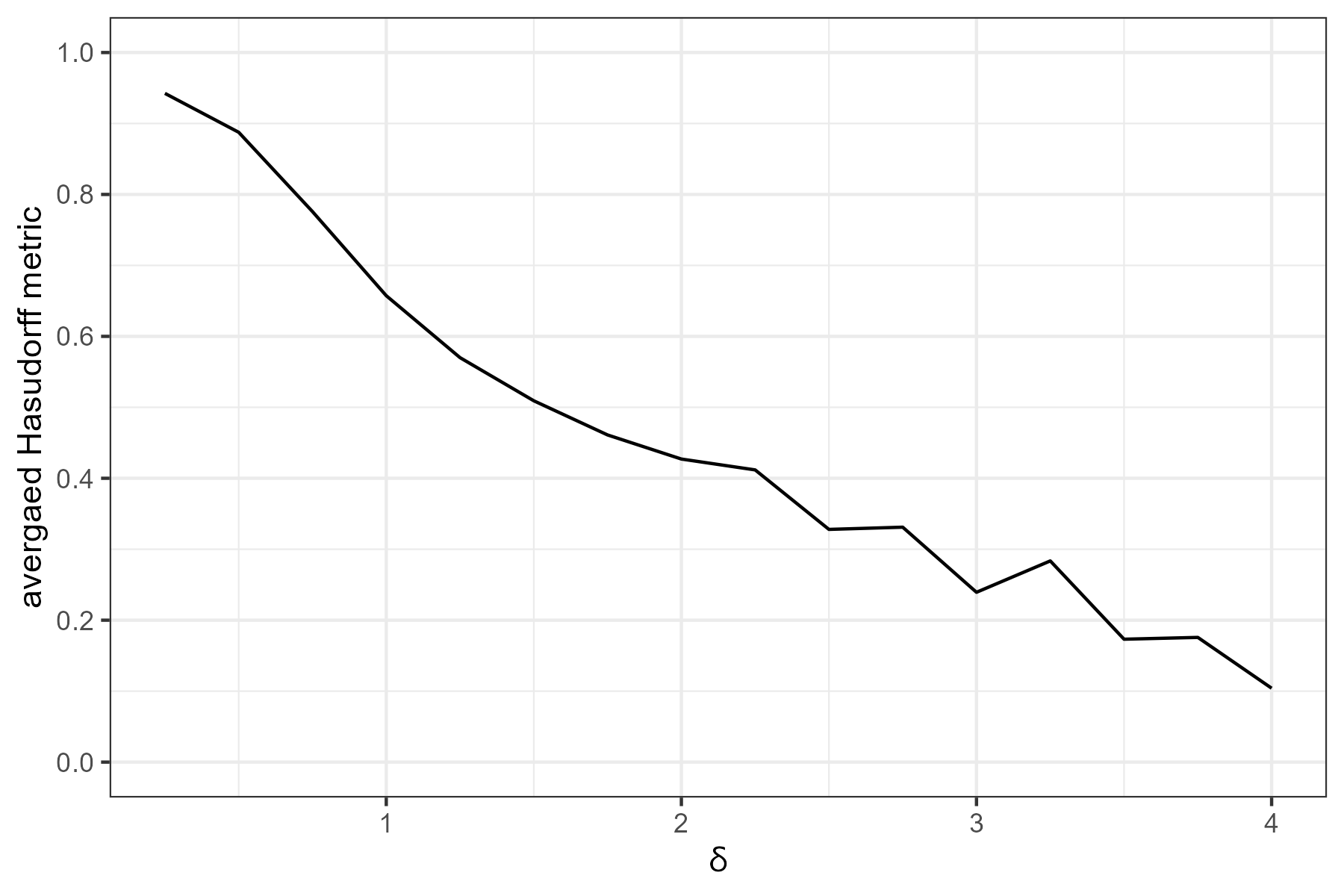}
    \caption{Scaling signal-to-noise ratio via $\delta$ and resulting\\ Hausdorff metric. \label{fig: mu_haus}} 
  \end{subfigure} 
      \caption{}
\end{figure}





\begin{figure}[h] {
\centering
\includegraphics[width=1\textwidth]{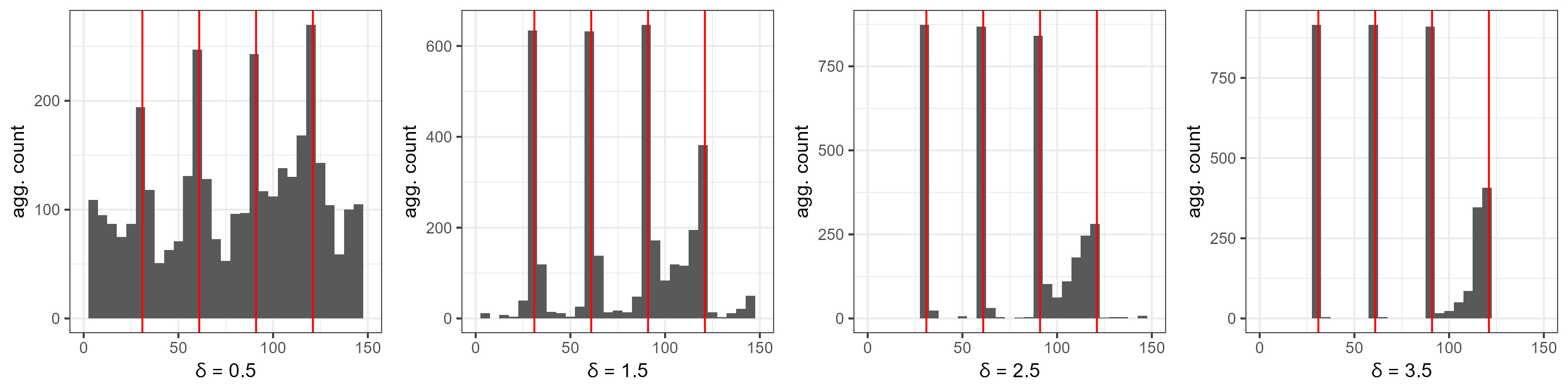}
\caption{Histogram of estimated change points over all Monte Carlo runs for selected values of $\delta$. \label{fig: mu_hist}} }
\end{figure}

There are two main insight obtained from this experiment. First, the approach yields the desired result in that both measures, F1-score and Hausdorff metric, improve for increasing signal-to-noise ratio $\delta$ with reliable estimation after $\delta\approx2.5$. The second, and much more valuable insight at this stage, is best portrayed by the histogram of change point locations over all Monte-Carlo simulations. The proposed algorithm seems to have a tendency in selecting change points in close proximity to each other for a given $K$ and struggles to recover the last change point -- even for larger values of $\delta$. To give a concrete example, it tends to choose $\mathbf{\hat{\tau}}=(31,61,90,91)$ or something similar for a given $K=4$ instead of the actual vector $\mathbf{\tau}$ and it seemingly does so consistently. This emphasizes the need of post-treatment of suggested candidates of change point locations, i.e., via clustering, before matching a given $K$. Furthermore, this behavior shows consistently throughout all experiments although we only include the histogram here.  

\subsubsection{Experiment A2 - Scaling $L$}

Again, set $K=4$ with the mean sequence of a simple staircase form with difference in mean $\delta=2$ at the change points, so $\mu_{\tau_k}-\mu_{\tau_k-1}=2$ for $k\in\{1,2,3,4\}$. We take $\mu_{\tau_0}=0$ for the artificial change point $\tau_0=1$ so it should not be detected. The choice of $\delta$ is hereby taken from the previous experiment indicating good detection for a minimal distance $L=30$ yet with potential for showing effect in either direction. Scaling now the distance between change points $L$ taking values from $15$ to $65$ with step-size $\triangle L=5$, we therefore have change points at $\tau=(L+1, 2L+1, 3L+1, 4L+1)$ and sequence length $n=5L$.

\begin{figure}[h!] 
  \begin{subfigure}{0.5\linewidth}
    \centering
    \includegraphics[width=0.7\textwidth]{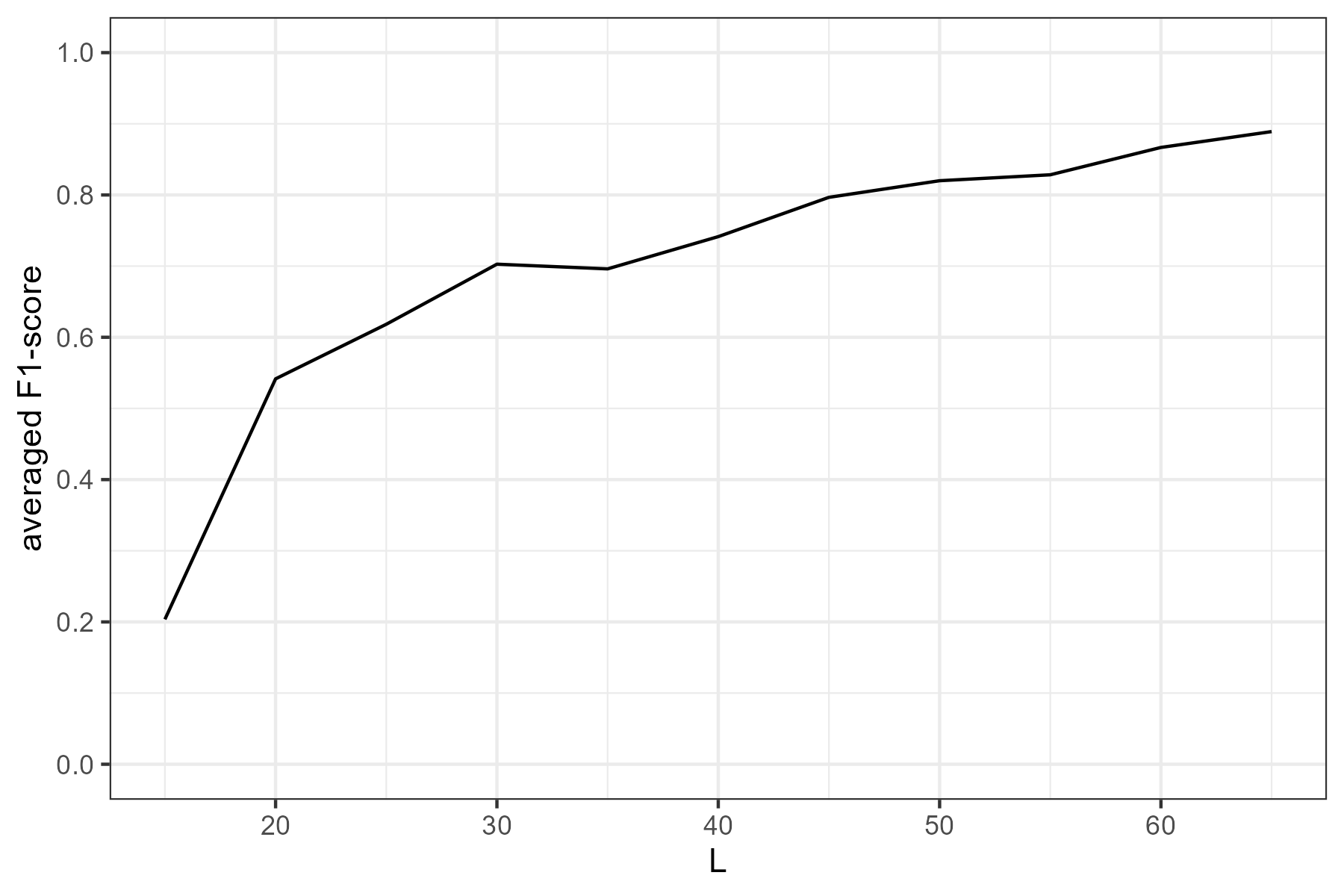}
    \caption{Scaling signal-to-noise ratio via $L$ and resulting\\
    F1-score. \label{fig: L_f1}} 
    \end{subfigure}
  \begin{subfigure}{0.5\linewidth}
    \centering
    \includegraphics[width=0.7\textwidth]{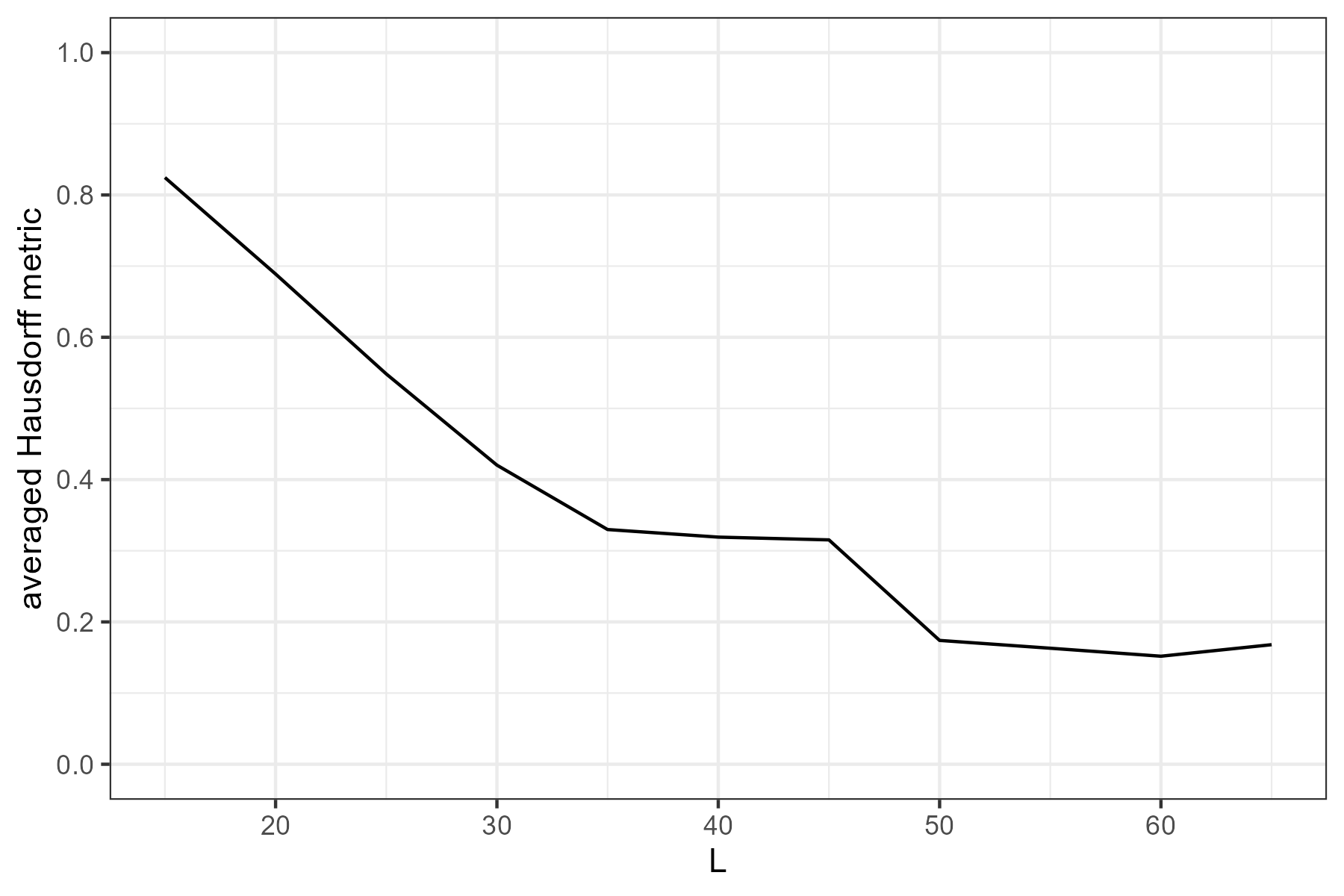}
    \caption{\textbf Scaling signal-to-noise ratio via $L$ and resulting\\ Hausdorff metric. \label{fig: L_haus}} 
  \end{subfigure} 
      \caption{}
\end{figure}




Again, w observe the desired behavior in improving performance measures for increasing $L$ tempering off for larger values. Accordingly, we may conclude the generally desired behavior of the approach in both metrics of interest for increasing signal strength $S$ even for this more pilot implementation.

\subsection{Experiment B - Unknown Number of Change Points}

For each Monte-Carlo simulation, the regularization penalty $\lambda$ for the COMBSS algorithm is chosen via both of the described methods. The discrepancy principle and the confidence bound hereby start at $\lambda=0$ and then increase with a step size $\triangle\lambda=0.005$ until the respective conditions are surpassed. The discrepancy principle then picks whichever $\lambda$ is closer to the expected error and the confidence bound chooses the previous penalty value $\lambda$ before the threshold is exceeded. Both estimate $\mathbf{\hat{\tau}}$ for their respective choice of~$\lambda$.

\subsubsection{Experiment B1 - Scaling $\delta^2$}

Set $n=100$ with a total of $3$ change points at $\tau=(26,51,76)$, so minimum distance $L=25$. The mean sequence has a simple staircase form with the difference of the mean at the change points $\mu_{\tau_k}-\mu_{\tau_k-1}$ for $k\in\{1,2,3\}$, so $\delta$, ranging from $1$ to $4$ with step-size $\triangle\delta=0.5$. The artificial change point is set to $\mu_{\tau_0}=0$, so it should not be detected. The necessary condition for detection derived in \cite{arias2005near} is at$\delta_\mathrm{crit}\geq\sqrt{\frac{2\log(100)}{25}}\approx0.61$ and is exceeded after the first step.

\begin{figure}[h!] 
  \begin{subfigure}{0.5\linewidth}
    \centering
    \includegraphics[width=0.7\textwidth]{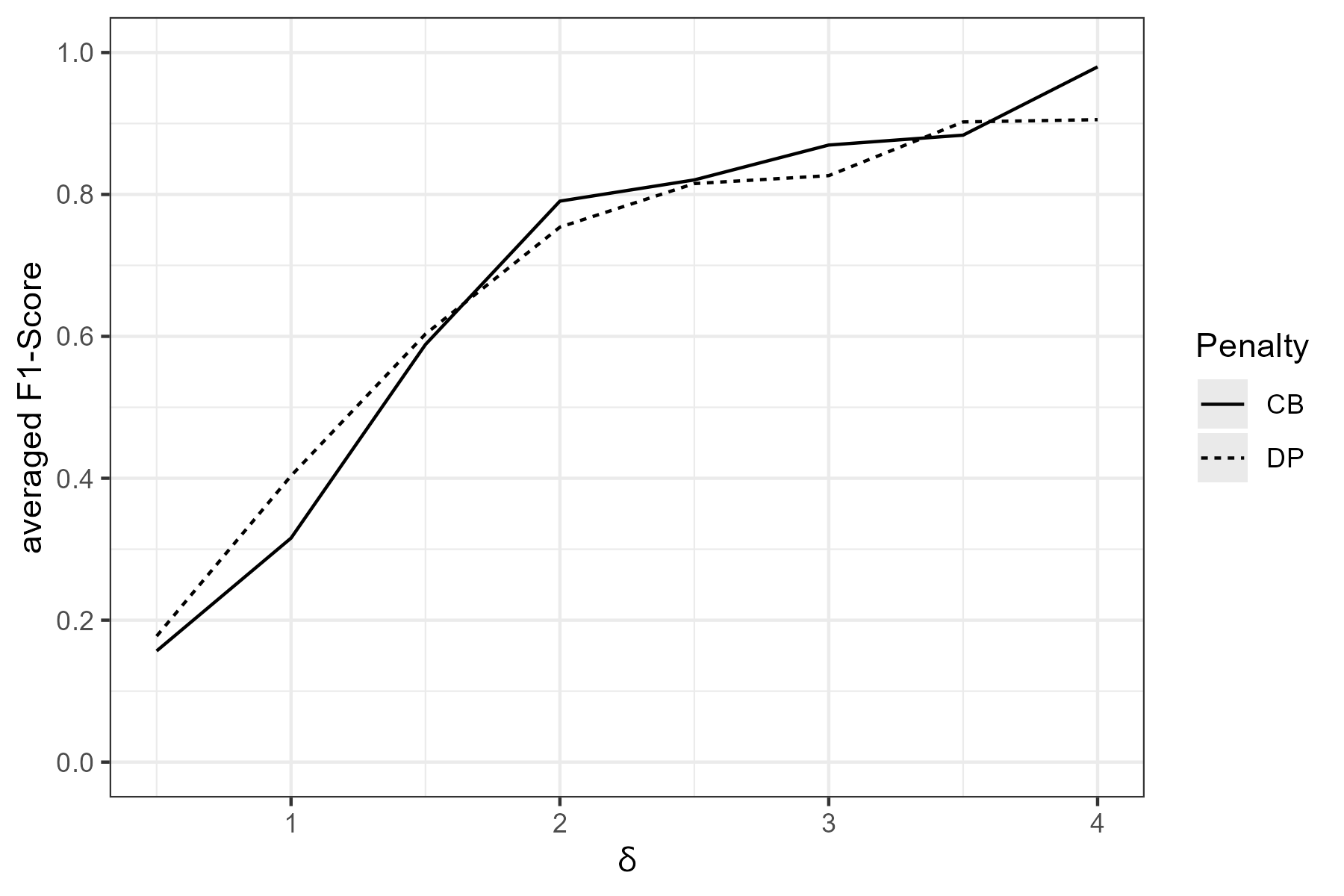}
    \caption{Scaling signal-to-noise ratio via $\delta$ and resulting\\ F1-score for both penalty choice approaches. \label{fig: mu_f1_DPCB}}
    \end{subfigure}
  \begin{subfigure}{0.5\linewidth}
    \centering
    \includegraphics[width=0.7\textwidth]{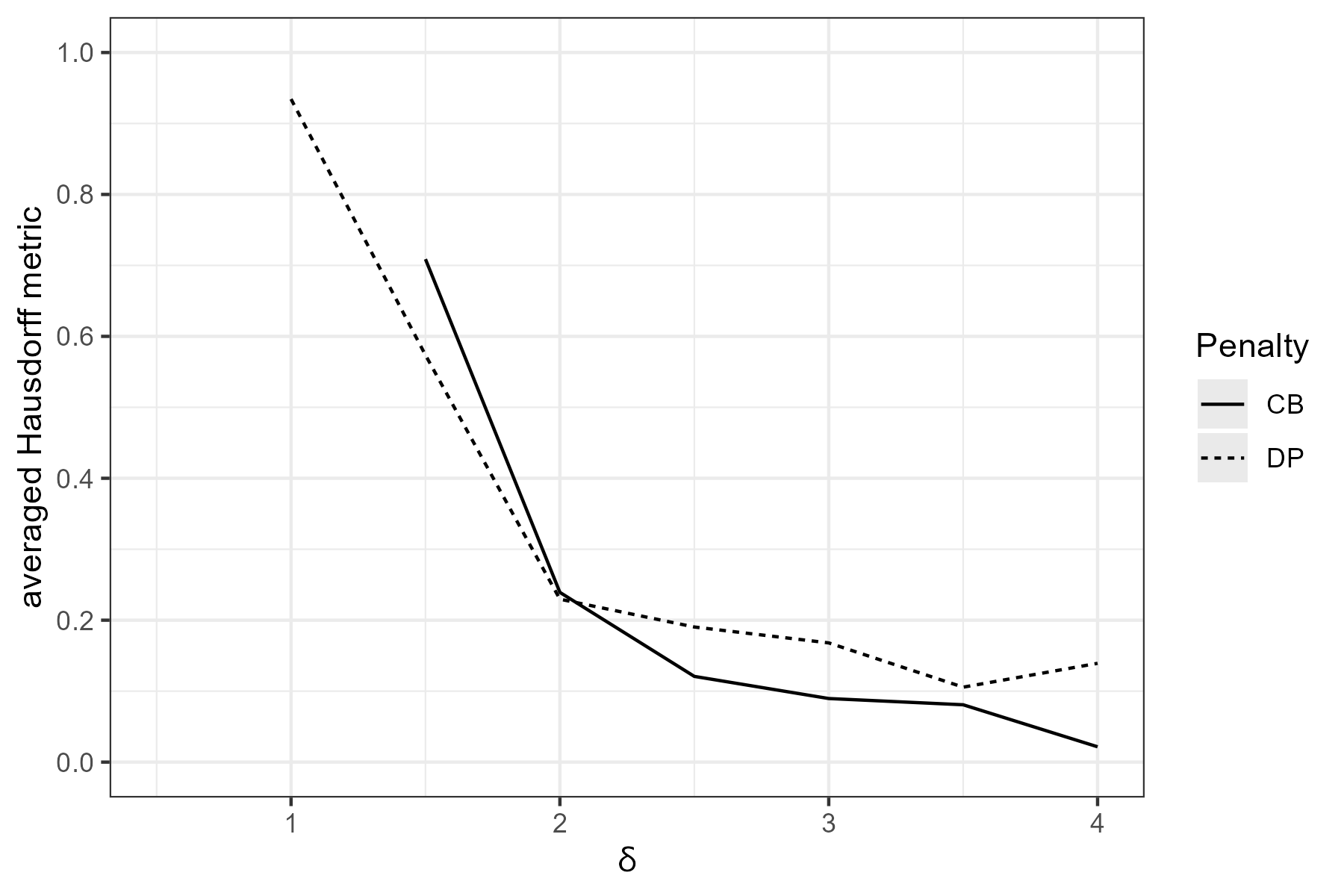}
    \caption{Scaling signal-to-noise ratio via $\delta$ and resulting\\ Hausdorff metric for both penalty choice approaches. \label{fig: mu_haus_DPCB}} 
  \end{subfigure} 
      \caption{}
\end{figure}



We observe the desired behavior for both curves, similar to the previous experiments. The confidence bound seems to provide a slightly better choice of $\lambda$ for larger values of $\delta$, but further investigation is needed.  

\subsubsection{Experiment B2 - Scaling $L$}

Again, set $K=3$ with the mean sequence of a simple staircase form with difference in mean $\delta=2$ at the change points $\mu_{\tau_k}-\mu_{\tau_k-1}=1$ for $k\in\{1,2,3\}$ and $\mu_{\tau_0}=0$ for the artificial change point $\tau_0=1$ so it should not be detected. $\delta$ is hereby chosen for the same reasons as in experiment A2 to provide range for chnage in performance in both directions. Scaling now the distance between change points $L$ taking values from $10$ to $50$ with step-size $\triangle L=5$, we therefore have change points at $\tau=(L+1, 2L+1, 3L+1)$ and sequence length $n=4L$.

\begin{figure}[h!] 
  \begin{subfigure}{0.5\linewidth}
    \centering
    \includegraphics[width=0.7\textwidth]{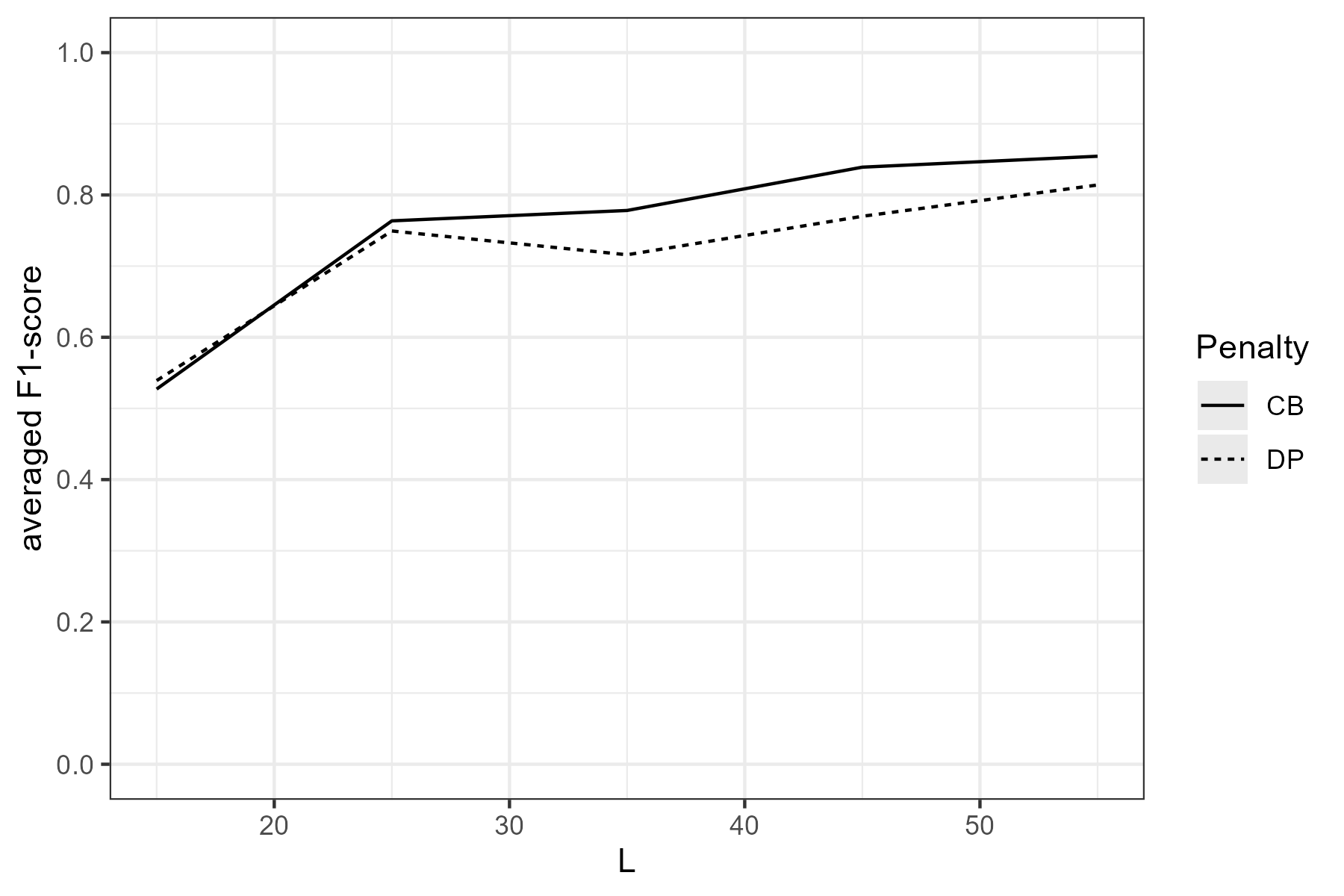}
    \caption{Scaling signal-to-noise ratio via $L$ and resulting\\ F1-score for both penalty choice approaches. \label{fig: L_f1_DPCB}}
    \end{subfigure}
  \begin{subfigure}{0.5\linewidth}
    \centering
    \includegraphics[width=0.7\textwidth]{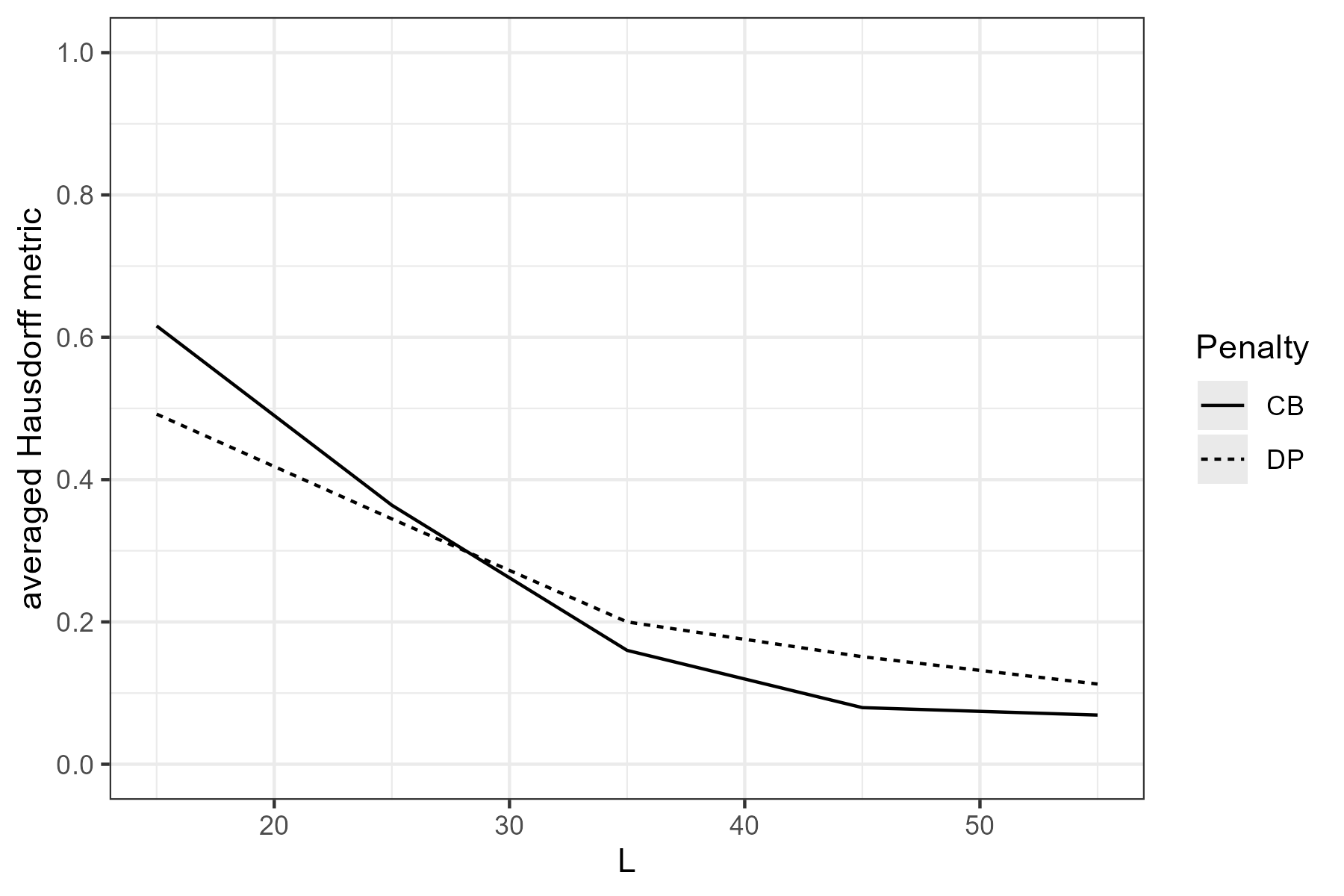}
    \caption{Scaling signal-to-noise ratio via $L$ and resulting\\ Hausdorff metric for both penalty choice approaches. \label{fig: L_haus_DPCB}}
  \end{subfigure} 
      \caption{}
\end{figure}



Similar to before, we observe improving performance in both metrics for increasing minimal distance $L$ between change points. There does not seem to be a major difference in the two choices of penalty value $\lambda$. Both methods seem viable with good performance and should be included in follow up studies.

\section{Discussion, Conclusions and Outlook}
\label{sec:conc}

The presented work is a proof of concept to show that there is great value to the approach utilizing COMBSS for candidates of change points in the normal mean change point model. This approach here still has plenty of room for improvement in implementation, yet the results provide a valuable basis for further investigation. While the experiments in the work at hand do not include a direct comparison to current Lasso approaches, results in \cite{moka2024combss} suggest for COMBSS competitive results in the more general case. Additionally, a real world application is highly desirable but needs to be subject for future investigations.\\
The main insights currently are the computational cost and the spurious change points estimated. The term spurious change points as mentioned in \cite{pilliat2023optimal} hereby refers to the tendency to detect change points in close proximity to each other. In practice, they can simply be aggregated into a single change point in post treatment whenever they are in too close proximity to each other. This problem mainly influenced the first two supervised experiments while the unsupervised experiments simply gave too large of a number of change points without this treatment, however, neither of the metrics were influenced by that in a meaningful way. For the F1-score, the tolerance accounted for that and in the Hausdorff metric only the closes change point is of interest. A possible line of future work arising from this problem could be to investigate 
the vast toolbox of linear regression for testing significance of supposed spurious change points. 
Other future lines of work will lie in exploring advances in adjacent topics. One such, is in investigating the discrepancy principle for choice of regularisation penalty under the sophisticated results in \cite{blanchard2012discrepancy}. The artificial design matrix is of trace type and can therefore be fit to their results for better statistical guarantees. 
Further, great interest lies in the gradient of the COMBSS algorithm given its specific choice of $X$. It promises high potential for improving computational efficiency and in combination with the approach of \cite{niu2016multiple} in utilizing symmetry by connecting the ends of a data sequence, it may help addressing the current issue of the neglected last change point.\\
\\
To conclude, utilizing the key concept in COMBSS in continuous relaxation of the best subset selection problem for modern gradient based optimization yields a novel, promising and valuable line of work for simultaneous offline multiple change point detection in the normal mean multiple change point model. It offers a large variety of directions for detailed investigations, optimizations and adaptations. With ongoing research on COMBSS, both fields may very much profit from each other sharing their advances. With the addition of COMBSS to the vast topic of change point detection has a new, additional connection to popular research in machine learning, further opening it up to interested researchers.

\bibliographystyle{unsrtnat}
\bibliography{ref}  

\begin{thebibliography}{32}
\providecommand{\natexlab}[1]{#1}
\providecommand{\url}[1]{\texttt{#1}}
\expandafter\ifx\csname urlstyle\endcsname\relax
  \providecommand{\doi}[1]{doi: #1}\else
  \providecommand{\doi}{doi: \begingroup \urlstyle{rm}\Url}\fi

\bibitem[Page(1954)]{page1954continuous}
Ewan~S Page.
\newblock Continuous inspection schemes.
\newblock \emph{Biometrika}, 41\penalty0 (1/2):\penalty0 100--115, 1954.

\bibitem[Chernoff and Zacks(1964)]{chernoff1964estimating}
Herman Chernoff and Shelemyahu Zacks.
\newblock Estimating the current mean of a normal distribution which is
  subjected to changes in time.
\newblock \emph{The Annals of Mathematical Statistics}, 35\penalty0
  (3):\penalty0 999--1018, 1964.

\bibitem[Lorden(1971)]{lorden1971procedures}
Gary Lorden.
\newblock Procedures for reacting to a change in distribution.
\newblock \emph{The Annals of Mathematical Statistics}, 42\penalty0
  (6):\penalty0 1897--1908, 1971.

\bibitem[Sen and Srivastava(1975)]{sen1975tests}
Ashish Sen and Muni~S Srivastava.
\newblock On tests for detecting change in mean.
\newblock \emph{The Annals of Statistics}, 3\penalty0 (1):\penalty0 98--108,
  1975.

\bibitem[Niu et~al.(2016)Niu, Hao, and Zhang]{niu2016multiple}
Yue~S Niu, Ning Hao, and Heping Zhang.
\newblock Multiple change-point detection: a selective overview.
\newblock \emph{Statistical Science}, 31\penalty0 (4):\penalty0 611--623, 2016.

\bibitem[Truong et~al.(2020)Truong, Oudre, and Vayatis]{truong2020selective}
Charles Truong, Laurent Oudre, and Nicolas Vayatis.
\newblock Selective review of offline change point detection methods.
\newblock \emph{Signal Processing}, 167:\penalty0 107299, 2020.

\bibitem[Ma et~al.(2020)Ma, Grant, and Sofronov]{Ma2020}
Lijing Ma, {Andrew J.} Grant, and Georgy Sofronov.
\newblock Multiple change point detection and validation in autoregressive time
  series data.
\newblock \emph{Statistical Papers}, 61\penalty0 (4):\penalty0 1507--1528,
  2020.
\newblock ISSN 0932-5026.

\bibitem[Moka et~al.(2024)Moka, Liquet, Zhu, and Muller]{moka2024combss}
Sarat Moka, Benoit Liquet, Houying Zhu, and Samuel Muller.
\newblock Combss: best subset selection via continuous optimization.
\newblock \emph{Statistics and Computing}, 34\penalty0 (2):\penalty0 75, 2024.

\bibitem[Huang et~al.(2005)Huang, Wu, Lizardi, and Zhao]{huang2005detection}
Tao Huang, Baolin Wu, Paul Lizardi, and Hongyu Zhao.
\newblock Detection of dna copy number alterations using penalized least
  squares regression.
\newblock \emph{Bioinformatics}, 21\penalty0 (20):\penalty0 3811--3817, 2005.

\bibitem[Rinaldo(2009)]{rinaldo2009properties}
Alessandro Rinaldo.
\newblock Properties and refinements of the fused lasso.
\newblock \emph{The Annals of Statistics}, 37\penalty0 (5B):\penalty0
  2922--2952, 2009.

\bibitem[Qian and Jia(2016)]{qian2016stepwise}
Junyang Qian and Jinzhu Jia.
\newblock On stepwise pattern recovery of the fused lasso.
\newblock \emph{Computational Statistics \& Data Analysis}, 94:\penalty0
  221--237, 2016.

\bibitem[Arias-Castro et~al.(2005)Arias-Castro, Donoho, and Huo]{arias2005near}
Ery Arias-Castro, David~L Donoho, and Xiaoming Huo.
\newblock Near-optimal detection of geometric objects by fast multiscale
  methods.
\newblock \emph{IEEE Transactions on Information Theory}, 51\penalty0
  (7):\penalty0 2402--2425, 2005.

\bibitem[Müller and Welsh(2010)]{mueller2010selection}
Samuel Müller and Alan~H. Welsh.
\newblock On model selection curves.
\newblock \emph{International Statistical Review}, 78\penalty0 (2):\penalty0
  240--256, 2010.

\bibitem[Hui et~al.(2017)Hui, M{\"u}ller, and Welsh]{hui2017joint}
Francis~KC Hui, Samuel M{\"u}ller, and AH~Welsh.
\newblock Joint selection in mixed models using regularized pql.
\newblock \emph{Journal of the American Statistical Association}, 112\penalty0
  (519):\penalty0 1323--1333, 2017.

\bibitem[Harchaoui and L{\'e}vy-Leduc(2010)]{harchaoui2010multiple}
Za{\i}d Harchaoui and C{\'e}line L{\'e}vy-Leduc.
\newblock Multiple change-point estimation with a total variation penalty.
\newblock \emph{Journal of the American Statistical Association}, 105\penalty0
  (492):\penalty0 1480--1493, 2010.

\bibitem[Tibshirani and Wang(2008)]{tibshirani2008spatial}
Robert Tibshirani and Pei Wang.
\newblock Spatial smoothing and hot spot detection for cgh data using the fused
  lasso.
\newblock \emph{Biostatistics}, 9\penalty0 (1):\penalty0 18--29, 2008.

\bibitem[Zhu et~al.(2020)Zhu, Wen, Zhu, Zhang, and Wang]{zhu2020polynomial}
Junxian Zhu, Canhong Wen, Jin Zhu, Heping Zhang, and Xueqin Wang.
\newblock A polynomial algorithm for best-subset selection problem.
\newblock \emph{Proceedings of the National Academy of Sciences}, 117\penalty0
  (52):\penalty0 33117--33123, 2020.

\bibitem[Hazimeh and Mazumder(2020)]{hazimeh2020fast}
Hussein Hazimeh and Rahul Mazumder.
\newblock Fast best subset selection: Coordinate descent and local
  combinatorial optimization algorithms.
\newblock \emph{Operations Research}, 68\penalty0 (5):\penalty0 1517--1537,
  2020.

\bibitem[Friedman et~al.(2007)Friedman, Hastie, H{\"o}fling, and
  Tibshirani]{friedman2007pathwise}
Jerome Friedman, Trevor Hastie, Holger H{\"o}fling, and Robert Tibshirani.
\newblock Pathwise coordinate optimization.
\newblock \emph{The Annals of Appplied Statistics}, 1\penalty0 (2):\penalty0
  302--332, 2007.

\bibitem[Wang et~al.(2020)Wang, Yu, and Rinaldo]{wang2020univariate}
Daren Wang, Yi~Yu, and Alessandro Rinaldo.
\newblock Univariate mean change point detection: Penalization, {CUSUM} and
  optimality.
\newblock \emph{Electronic Journal of Statistics}, 14:\penalty0 1917--1961,
  2020.

\bibitem[Verzelen et~al.(2023)Verzelen, Fromont, Lerasle, and
  Reynaud-Bouret]{verzelen2023optimal}
Nicolas Verzelen, Magalie Fromont, Matthieu Lerasle, and Patricia
  Reynaud-Bouret.
\newblock Optimal change-point detection and localization.
\newblock \emph{The Annals of Statistics}, 51\penalty0 (4):\penalty0
  1586--1610, 2023.

\bibitem[Natarajan(1995)]{natarajan1995sparse}
Balas~Kausik Natarajan.
\newblock Sparse approximate solutions to linear systems.
\newblock \emph{SIAM journal on computing}, 24\penalty0 (2):\penalty0 227--234,
  1995.

\bibitem[Bertsimas et~al.(2016)Bertsimas, King, and
  Mazumder]{bertsimas2016best}
Dimitris Bertsimas, Angela King, and Rahul Mazumder.
\newblock Best subset selection via a modern optimization lens.
\newblock \emph{The Annals of Statistics}, 44\penalty0 (2):\penalty0 813--852,
  2016.

\bibitem[Furnival and Wilson(2000)]{furnival2000regressions}
George~M Furnival and Robert~W Wilson.
\newblock Regressions by leaps and bounds.
\newblock \emph{Technometrics}, 42\penalty0 (1):\penalty0 69--79, 2000.

\bibitem[Richter(2021)]{richter2021inverse}
Mathias Richter.
\newblock \emph{Inverse Problems: Basics, Theory and Applications in
  Geophysics}.
\newblock Springer Nature, 2021.

\bibitem[Blanchard and Math{\'e}(2012)]{blanchard2012discrepancy}
Gilles Blanchard and Peter Math{\'e}.
\newblock Discrepancy principle for statistical inverse problems with
  application to conjugate gradient iteration.
\newblock \emph{Inverse problems}, 28\penalty0 (11):\penalty0 115011, 2012.

\bibitem[Tian and Takane(2005)]{tian2005schur}
Yongge Tian and Yoshio Takane.
\newblock Schur complements and banachiewicz-schur forms.
\newblock \emph{The Electronic Journal of Linear Algebra}, 13:\penalty0
  405--418, 2005.

\bibitem[Woodbury(1950)]{woodbury1950inverting}
Max~A Woodbury.
\newblock \emph{Inverting modified matrices}.
\newblock Department of Statistics, Princeton University, 1950.

\bibitem[Lee(2011)]{lee2011tridiagonal}
WT~Lee.
\newblock Tridiagonal matrices: Thomas algorithm.
\newblock \emph{MS6021, Scientific Computation, University of Limerick}, 2011.

\bibitem[Higham(2002)]{higham2002accuracy}
Nicholas~J Higham.
\newblock \emph{Accuracy and Stability of Numerical Algorithms}.
\newblock SIAM, 2002.

\bibitem[Aminikhanghahi and Cook(2017)]{aminikhanghahi2017survey}
Samaneh Aminikhanghahi and Diane~J Cook.
\newblock A survey of methods for time series change point detection.
\newblock \emph{Knowledge and information systems}, 51\penalty0 (2):\penalty0
  339--367, 2017.

\bibitem[Pilliat et~al.(2023)Pilliat, Carpentier, and
  Verzelen]{pilliat2023optimal}
Emmanuel Pilliat, Alexandra Carpentier, and Nicolas Verzelen.
\newblock Optimal multiple change-point detection for high-dimensional data.
\newblock \emph{Electronic Journal of Statistics}, 17\penalty0 (1):\penalty0
  1240--1315, 2023.

\end{thebibliography}
\end{document}